%% file: decoupht.tex
\documentclass[12pt]{article}
\usepackage{epsfig}

\newcommand\pubnumber{IFT-UAM/CSIC 01-03, FTUAM 01/02, KA-TP-3-2001}
\newcommand\pubdate{\today}
\newcommand\hepnumber{hep-ph/0102169}

\def\csumb{$^a$ Santa Cruz Institute for Particle Physics, University
of California. \\
$^b$ Departamento de F\'{\i}sica Te\'{o}rica, Universidad Aut\'{o}noma
de Madrid. \\
$^c$ Theoretical Physics Department, Fermi National Accelerator Laboratory.\\
$^d$ Institut f\"ur Theoretische Physik, Universit\"at Karlsruhe.\\  
$^e$ Department of Physics, University of Michigan. }
\def\support{\footnote{e-mails: haber@scipp.ucsc.edu, herrero@delta.ft.uam.es,
logan@fnal.gov,\\ siannah@particle.uni-karlsruhe.de, srigolin@umich.edu, 
temes@delta.ft.uam.es}}

\def\Title#1{\begin{center} {\Large\bf #1 } \end{center}}
\def\Author#1{\begin{center}{ \sc #1} \end{center}}
\def\Address#1{\begin{center}{ \it #1} \end{center}}

\newcommand\pubblock{\rightline{\begin{tabular}{l} \pubnumber\\
         \pubdate \\ \hepnumber \end{tabular}}}
\newenvironment{Abstract}{\begin{quotation}  }{\end{quotation}}
\newenvironment{Presented}{\begin{quotation} \begin{center} 
             Presented at the\end{center}
      \begin{center}\begin{large}}{\end{large}\end{center} \end{quotation}}
\def\Acknowledgments{\bigskip  \bigskip \begin{center}
          \large\bf Acknowledgments\end{center}}

\makeatletter
\def\section{\@startsection{section}{0}{\z@}{5.5ex plus .5ex minus
 1.5ex}{2.3ex plus .2ex}{\large\bf}}
\def\subsection{\@startsection{subsection}{1}{\z@}{3.5ex plus .5ex minus
 1.5ex}{1.3ex plus .2ex}{\normalsize\bf}}
\def\subsubsection{\@startsection{subsubsection}{2}{\z@}{-3.5ex plus
-1ex minus  -.2ex}{2.3ex plus .2ex}{\normalsize\sl}}

\renewcommand{\@makecaption}[2]{%
   \vskip 10pt
   \setbox\@tempboxa\hbox{\small #1: #2}
   \ifdim \wd\@tempboxa >\hsize     % IF longer than one line:
       \small #1: #2\par          %   THEN set as ordinary paragraph.
     \else                        %   ELSE  center.
       \hbox to\hsize{\hfil\box\@tempboxa\hfil}
   \fi}

 \def\citenum#1{{\def\@cite##1##2{##1}\cite{#1}}}
\newcount\@tempcntc
\def\@citex[#1]#2{\if@filesw\immediate\write\@auxout{\string\citation{#2}}\fi
  \@tempcnta\z@\@tempcntb\m@ne\def\@citea{}\@cite{\@for\@citeb:=#2\do
    {\@ifundefined
       {b@\@citeb}{\@citeo\@tempcntb\m@ne\@citea\def\@citea{,}{\bf ?}\@warning
       {Citation `\@citeb' on page \thepage \space undefined}}%
    {\setbox\z@\hbox{\global\@tempcntc0\csname b@\@citeb\endcsname\relax}%
     \ifnum\@tempcntc=\z@ \@citeo\@tempcntb\m@ne
       \@citea\def\@citea{,}\hbox{\csname b@\@citeb\endcsname}%
     \else
      \advance\@tempcntb\@ne
      \ifnum\@tempcntb=\@tempcntc
      \else\advance\@tempcntb\m@ne\@citeo
      \@tempcnta\@tempcntc\@tempcntb\@tempcntc\fi\fi}}\@citeo}{#1}}
\def\@citeo{\ifnum\@tempcnta>\@tempcntb\else\@citea\def\@citea{,}%
  \ifnum\@tempcnta=\@tempcntb\the\@tempcnta\else
  {\advance\@tempcnta\@ne\ifnum\@tempcnta=\@tempcntb \else\def\@citea{--}\fi
    \advance\@tempcnta\m@ne\the\@tempcnta\@citea\the\@tempcntb}\fi\fi}
\makeatother

\input econf2.tex
\begin{document}
\begin{titlepage}
\begin{flushright}
\pubblock
\end{flushright}

\vfill
\def\thefootnote{\fnsymbol{footnote}}
\Title{Decoupling Properties of MSSM particles \\[5pt] in Higgs and 
Top Decays}
\vfill
\Author{ H. E. Haber$^a$, \underline {M. J. Herrero}$^{b}$, 
H. E. Logan$^c$,
\\[6pt]
S. Pe\~naranda$^d$, S. Rigolin$^e$
and D. Temes$^b$\support}
\Address{\csumb}
\vfill
\begin{Abstract}
We study the supersymmetric (SUSY) QCD radiative corrections, at the one-loop
level, to $h^0$, $H^{\pm}$ and $t$ quark decays, in the 
context of the Minimal Supersymmetric Standard Model (MSSM) and in the 
decoupling limit. The decoupling behaviour of the various MSSM 
sectors is analyzed in some special cases, where some or all of the SUSY mass
parameters become large as compared to the electroweak scale. We show that in
the decoupling limit of both large SUSY mass parameters and large CP-odd Higgs 
mass,
the $\Gamma (h^0\rightarrow b \bar b) $ decay width approaches its Standard
Model value at one loop, with the onset of decoupling being delayed for large
$\tan\beta$ values. However, this decoupling does not occur if just the  
SUSY mass parameters are taken large. A similar interesting non-decoupling 
behaviour, also enhanced by $\tan\beta$, is found in the SUSY-QCD  
corrections to the
$\Gamma (H^+\rightarrow t \bar b) $ decay width at one loop. In contrast, 
the SUSY-QCD
corrections in the  $\Gamma (t\rightarrow W^+ b) $ decay width do decouple and
this decoupling is fast.
\end{Abstract}
\vfill
\begin{Presented}
5th International Symposium on Radiative Corrections \\ 
(RADCOR--2000) \\[4pt]
Carmel CA, USA, 11--15 September, 2000
\end{Presented}
\vfill
\end{titlepage}
\def\thefootnote{\arabic{footnote}}
\setcounter{footnote}{0}

\section{Introduction}
 The study of radiative corrections 
to Standard Model (SM) couplings may provide crucial 
clues in exploring 
new physics beyond the 
reach of present accelerators. In particular, suppose that a light Higgs boson, $h^0$,
 were discovered 
in the mass range predicted by the minimal supersymmetric extension of
the Standard Model (MSSM),
but supersymmetric (SUSY) particles were not found. Then, a precise 
measurement of Higgs couplings to SM particles, which are sensitive to 
radiative corrections, could provide
indirect information about the existence of SUSY in Nature and some
indication of the   
preferred region of the SUSY parameter space.  For example, one could
predict (in the context of the MSSM)
whether the data favored a SUSY spectrum below the 1 TeV energy scale.
Similar studies can be performed by considering alternative observables as,
for instance,  
the partial widths of top and Higgs decays into SM particles, and by comparing 
their predictions in the MSSM and the SM. 

In this comparison of the MSSM and SM predictions for observables involving SM particles 
in the external legs, it is interesting to consider some particular
limiting situations.
The first one is when the genuine SUSY spectrum is very heavy as compared to the 
electroweak scale, $M_{SUSY}\gg M_Z$, where $M_{SUSY}$ represents generically 
the masses of the SUSY particles. This situation corresponds to the decoupling 
of SUSY particles from the rest of the MSSM spectrum, namely, the SM particles 
and the MSSM Higgs sector containing $h^0$, $H^0$, $A^0$ and $H^\pm$.
 The second one is when the extra Higgs bosons, $H^0$, $A^0$ 
and $H^\pm$ are very heavy, but $h^0$ and the genuine SUSY particles are closer
to the electroweak scale. This decoupling limit can be reached by considering 
$M_A \gg M_Z$, where
$M_A$ is the mass of the CP-odd neutral Higgs boson of the MSSM.
In addition, there is the limiting situation where both $M_{SUSY}$ and $M_A$ are
large,  and the decoupling of all non-standard particles from the SM physics
is expected. This decoupling is known to occur in tree-level physics and in 
some one-loop physics. In
particular, the tree-level couplings of  $h^0$ to fermion pairs and gauge bosons
tend to their SM values if $M_A \gg M_Z$~\cite{decoupling}.
As a consequence of this decoupling, distinguishing
the lightest MSSM Higgs boson in the large $M_A$
limit from the Higgs boson of the SM will be very
difficult.

Formally, the decoupling of all non-standard MSSM particles implies that
in the effective low-energy theory, all observables involving SM particles in
the external legs tend
to their SM values in the limit of large SUSY masses and large $M_A$.
It has been shown that all the genuine SUSY particles in the MSSM and 
the heavy Higgs bosons $H^0$, $A^0$ and $H^\pm$, decouple at one-loop order
from the low-energy electroweak gauge boson physics~\cite{dhp}.
In particular, the contributions of the SUSY particles
to low-energy processes either fall as inverse powers of the SUSY
mass parameters or can be absorbed into counterterms for the tree-level
couplings of the low-energy theory and, therefore, they decouple in the same
spirit as established in the {\it Appelquist-Carazzone Theorem}~\cite{ac}. 
 As a result,
the radiative corrections involving SUSY particles go to zero
in the asymptotic large SUSY mass limit.

Our purpose here is to determine the decoupling behaviour in the previous limiting
situations of several observables, including radiative corrections  
at one-loop, with the hope that for some of them either 
the decoupling does not occur totally or, in case it occurs, it proceeds 
slowly
such that there may remain significant signals of new physics beyond SM, even
for a heavy SUSY spectra.     

In this paper we focus on the partial widths of $h^0\rightarrow b \bar b$,
$H^+\rightarrow t \bar b$ and $t \rightarrow W^+ b$ decays, with special
emphasis on the first one, whose corresponding branching ratio will be crucial
for the experimental Higgs boson searches at the upcoming Tevatron
Run 2~\cite{cmwpheno,habtev}. We study the MSSM radiative corrections to these
observables at the one-loop level
and to leading order
in $\alpha_s$, and we analyze in detail their behavior in the
previously mentioned  
decoupling limits. These corrections are due to the SUSY-QCD (SQCD) sector and 
arise from
gluinos and third generation-squark
exchange.  Because of the dependence on the strong
coupling constant, these are expected to be the most significant
one-loop MSSM contributions over much of the MSSM parameter space.
We will show that in the limit of large
$M_A$ (in this limit one also has $M_{H^0},M_{H^\pm} \gg M_Z$)
and large sbottom and gluino masses ($M_{\tilde b_i},M_{\tilde g} \gg M_Z$),
the SM expression for the $h^0 \rightarrow b\bar b$ one-loop 
partial width is recovered~\cite{Haber:2001kq}.
That is, the SQCD corrections to the $\Gamma(h^0 \rightarrow b \bar b)$
partial width decouple in the limit of large SUSY masses and large $M_A$.
In particular, we examine the case of large $\tan\beta$, for which
the SQCD corrections are enhanced.
This enhancement can delay the onset of
decoupling and give rise to a significant one-loop correction, even for
moderate to large values of the SUSY masses. This decoupling, however, does not
occur, if either $M_{SUSY}$ (characterizing a common mass scale for 
gluino and sbottom masses) or $M_A$ are kept fixed while 
the other is taken large. A similar non-decoupling phenomenon of the SQCD
corrections to one-loop when $M_A$ is fixed and the sbottom, stop and gluinos masses are
considered large 
is found in the $H^+ \rightarrow t\bar b$ decay~\cite{hpt}. 
The SQCD corrections to
one-loop in 
the $t\rightarrow W^+ b$ decay,
however, do decouple and this decoupling  proceeds fast.
We present here just a summary of the main results and refer the reader 
to refs.~\cite{Haber:2001kq,hpt} for more details.

%%%%%%%%%%%%%%%%%%%%%%%%%%%%%%%%%%%%%%%%%%%%%%   

\section{Decoupling  limit in the Higgs sector}

 The decoupling limit in the Higgs sector of the MSSM was
 first studied in ref.~\cite{decoupling}. In short, it is defined by considering
 the CP-odd Higgs mass much larger than the electroweak scale, $M_A \gg M_Z$,
 and leads to a particular spectrum in the Higgs sector 
 with very heavy $H^0$, $H^{\pm}$ and $A^0$ bosons, and a light $h^0$ boson.
 For a review of the MSSM Higgs sector, see ref.~\cite{HHG}.
  
 At tree level, if $M_A \gg M_Z$, the Higgs masses are, 

\vspace{0.1cm}
 \hspace{2.5cm}
 $ M_{H^o} \simeq M_{H^{\pm}} \simeq M_A \gg M_Z 
\,\,\,,\,\, M_{h^o} \simeq M_Z |cos2\beta|\,.$\\
That is, at tree-level there exists  a  CP-even Higgs, $h^0$, lighter than the $Z$ boson.
 
 Concerning the neutral Higgs couplings, their tree-level values in the 
 MSSM normalized 
 to SM couplings and for arbitrary $M_A$, are given in table~\ref{hcoup}. 
  
\begin{table}[hbt]
\renewcommand{\arraystretch}{1.5}
\begin{center}
\begin{tabular}{|lc||ccc|} \hline
\multicolumn{2}{|c||}{$\phi$} & $g_{\phi \bar tt}$ & $g_{\phi \bar bb}$ & 
 $g_{\phi VV}$ \\
\hline \hline
SM~ & $H$ & 1 & 1 & 1 \\ \hline
MSSM~ & $h^o$ & $ {\cos\alpha/\sin\beta}$ & $ {-\sin\alpha/\cos\beta}$ &
$ {\sin(\beta-\alpha)}$ \\
& $H^o$ & $\sin\alpha/\sin\beta$ & $\cos\alpha/\cos\beta$ &
$\cos(\beta-\alpha)$ \\
& $A^o$ & $ 1/\tan\beta$ & $\tan\beta$ & 0 \\ \hline
\end{tabular}
\renewcommand{\arraystretch}{1.2}
%\vspace{-1.2cm}
\caption{\label{hcoup} Higgs couplings in the MSSM normalized to SM 
couplings} 
\end{center}
\end{table}
\vspace{-0.5cm}

Notice that by expanding in inverse powers of $M_A$, we get:

\vspace{0.1cm}

\hspace{2cm} $ {\frac{\cos\alpha}{\sin\beta}}\simeq  1 + 
\cal{O}$$(M_Z^2/M_A^2)$$\,,\,$
$ {-\frac{\sin\alpha}{\cos\beta}}\simeq 1 + 
\cal{O}$$(M_Z^2/M_A^2)$

\vspace{0.3cm}

\hspace{3.5cm} $ {\sin(\beta - \alpha)} \simeq 1 + 
\cal{O}$$(M_Z^4/M_A^4)$.
\vspace{0.1cm}

Therefore, the $h^0$ tree-level couplings in the decoupling limit, 
$M_A \gg M_Z$,
tend to their SM values, as expected.

Beyond tree level, it has been shown~\cite{hmassdecoup} that, in this same decoupling 
limit, the Higgs masses keep a similar pattern as at tree level, that is, 
very heavy $H^0$, $H^{\pm}$ and $A^0$ bosons, and a light $h^0$ boson. The
particular values of their masses depend of course on the MSSM parameters, 
but for 
$M_A \gg M_Z$,  

\vspace{0.1cm}

\hspace{4cm}{  $M_{H^o} \simeq M_{H^{\pm}} \simeq M_A \gg M_Z\,,$}

\vspace{0.2cm}

\hspace{4.5cm}{ $M_{h^o} \le  130-135 \, GeV\,.$} 
\vspace{0.1cm}

In this work we will go beyond tree level and study    
  the decoupling behaviour of heavy  
 SUSY particles and heavy Higgses, at one-loop level,
  in Higgs bosons and top quark decays.  
  
%%%%%%%%%%%%%%%%%%%%%%%%%%%%%%%%%%%%%%%%%%%%%%%%%%%%
\section{Decoupling limit in the SUSY-QCD sector}  

The sbottom and stop mass matrices, in the MSSM, are given respectively by:

\vspace{-0.4cm}

$${\hat M}_{\tilde{b}}^2 =\left(\begin{array}{cc}  
 { {M_{{\scriptscriptstyle {\tilde Q}}}^{2}}+ m_{b}^{2} - 
 M_{Z}^{2} (\frac{1}{2}+Q_b   
s_{{\scriptscriptstyle W}}^{2}) \cos{2\beta}} & 
{m_{\scriptscriptstyle b}( {A_b}- {\mu} \tan{\beta})}  
\\ {m_{\scriptscriptstyle b}( {A_b}- {\mu} \tan{\beta})} &
{  {M_{{\scriptscriptstyle {\tilde D}}}^{2}} +    
m_{b}^{2} + M_{Z}^{2} Q_b s_{{\scriptscriptstyle W}}^{2} \cos{2\beta}}   
\end{array} \right) $$

and
   
$${\hat M}_{\tilde{t}}^2 =\left(\begin{array}{cc}  
 { {M_{{\scriptscriptstyle {\tilde Q}}}^{2}}+ m_{t}^{2} +
  M_{Z}^{2} (\frac{1}{2}-Q_t   
s_{{\scriptscriptstyle W}}^{2}) \cos{2\beta}} & 
{m_{\scriptscriptstyle t}( {A_t}- {\mu} \cot{\beta})}  
\\ {m_{\scriptscriptstyle t}( {A_t}- {\mu} \cot{\beta})} &
{  {M_{{\scriptscriptstyle {\tilde U}}}^{2}} +   
m_{t}^{2} + M_{Z}^{2} Q_t s_{{\scriptscriptstyle W}}^{2} \cos{2\beta}}  
\end{array} \right) $$

In order to get heavy squarks and heavy gluinos, we need to choose 
properly the soft SUSY breaking parameters and the $\mu$-parameter. Since here 
we are interested in the limiting situation where the whole 
SUSY spectrum is heavier than the electroweak scale, we have made
the following assumptions for the soft breaking squark mass parameters, trilinear
terms, $\mu$-parameter and gluino mass (see ref.~\cite{Haber:2001kq} for more 
details),

$$M_{SUSY} \sim M_{{\scriptscriptstyle {\tilde Q}}} 
\sim M_{{\scriptscriptstyle {\tilde D}}}
\sim M_{{\scriptscriptstyle {\tilde U}}}
\sim M_{\tilde g} 
\sim \mu  \sim A_{b} \sim A_{t} \gg M_Z, $$ 
where $M_{SUSY}$ represents generically a common SUSY large mass scale.

 Besides, we have considered two extreme cases, maximal and minimal mixing,  
 which, for the large mass limit we are studing, imply certain constraints 
 on the squark mass differences.
 Thus, given the generic mass matrix,   

\vspace{-0.3cm}

$${\hat M}^2_{\tilde q} \equiv\left(\begin{array}{cc}  
 { M_L^2} & 
{m_q X_q}  
\\ {m_q X_q} &
{M_R^2}   
\end{array} \right), $$
the two limiting cases are reached by choosing the relative size of 
$M_{L,R}$ and $X_q$ as follows,

%\vspace{0.5cm}
A.-Close to maximal mixing:
$\theta_{\tilde q} \sim \pm 45^{\circ}$
 
\vspace{-0.2cm}

$$|M_L^2-M_R^2|\ll m_q X_q \Rightarrow 
|M_{\tilde q_1}^2-M_{\tilde q_2}^2|\ll 
|M_{\tilde q_1}^2+M_{\tilde q_2}^2|$$

%\vspace{0.5cm}
B.-Close to minimal mixing:
 $\theta_{\tilde q} \sim 0^{\circ}$
 
\vspace{-0.2cm}

$$|M_L^2-M_R^2|\gg m_q X_q \Rightarrow 
|M_{\tilde q_1}^2-M_{\tilde q_2}^2|\sim
{\cal O}|M_{\tilde q_1}^2+M_{\tilde q_2}^2|$$
Here we have included the corresponding implications for the squark mass
differences.
%%%%%%%%%%%%%%%%%%%%%%%%%%%%%%%%%%%%%%%%%%%%%%%%%%
\section{SUSY-QCD corrections to $h^o \rightarrow \bar bb$ in the decoupling
limit}

In this section we study the SUSY-QCD corrections to the partial 
decay width $\Gamma(h^o \rightarrow \bar bb)$ at the one-loop level and to leading
order in perturbative QCD, that is ${\cal O}$($\alpha_S$). We will then explore
the decoupling behaviour of these corrections for large SUSY masses, $M_{SUSY}$,
and/or large $M_A$. Both numerical and analytical results will be 
presented~\cite{Haber:2001kq}.

For the $h^0$ mass range predicted by the MSSM, the decay channel
$h^o \rightarrow \bar bb$ is by far the dominant one (except in some
special regions of parameter space at large $\tan \beta$), and the precise 
value of its branching ratio 
will be crucial for the $h^0$ final experimental reach at the Tevatron.
  
Among the various contributions to this decay width, the QCD 
 corrections are known to be the dominant ones. At the one-loop level and to order 
$\alpha_S$ these can be written as,  
$$\Gamma_1(h^o \to b \bar b) \equiv \Gamma_0(h^o \to b \bar b)
        (1 + 2   {\Delta_{QCD}} + 2    {\Delta_{SQCD}}),$$
where, $\Gamma_0(h^o \to b \bar b)$ is the tree level width, $\Delta_{QCD}$ 
is the one-loop contribution from standard QCD  and, $\Delta_{SQCD}$ is the one-loop
contribution from the SUSY-QCD sector of the MSSM. The QCD correction, 
$\Delta_{QCD}$, gives a $\sim  50 \%$  reduction in the
 $\Gamma (h^o \rightarrow b \bar b)$ decay rate for $M_{h^0}$ in its MSSM 
 range~\cite{qcd} . 
  This correction has the same form in the MSSM as in the SM, so that
 it gives no information in distinguishing the MSSM from the SM. 
 The SQCD correction, $\Delta_{SQCD}$, was first computed in the
  on-shell scheme by using a
  diagrammatic approach in ref.~\cite{Dabelstein} and later studied 
in detail in~\cite{cjs}. The SQCD corrections to the $h^0b\bar b$
  coupling  were
  also computed in an effective Lagrangian approach in 
  ref.~\cite{cmwpheno}, using the SUSY contributions to
  the $b$-quark self energy~\cite{hrs-h-copw,pbmz}
  and neglecting terms suppressed by inverse powers of
  SUSY masses. The size of the 
 SQCD 
 correction, $\Delta_{SQCD}$, 
     and the QCD correction, $\Delta_{QCD}$, are comparable 
  for a wide window of the MSSM parameter space. In some regions of the MSSM parameter space, the SQCD corrections
  become so large that it is important to take into account higher-order
  corrections.  The two-loop SQCD corrections have been studied in a
  diagrammatic approach in
  ref.~\cite{hhw}. A higher-order analysis 
   has also been carried
  out in refs.~\cite{cgnw,ehkmy} by
resumming the leading $\tan\beta$ contributions to all orders of perturbation
theory and by using an effective Lagrangian approach. However, this
resummation is not important in our present work because
we are interested in the decoupling limit, in which the one-loop
corrections to the $h^0 b\bar b$ coupling are small. Thus, for the present
analysis we will just keep the one-loop corrections.
\begin{figure}[h]
\begin{center}
\epsfig{file=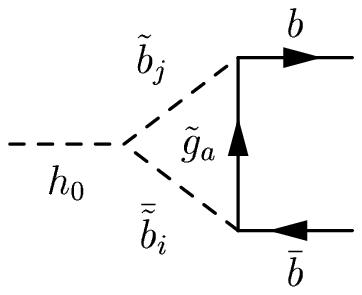}~~
\epsfig{file=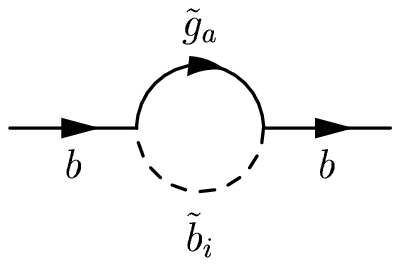}
\caption[0]{One-loop SUSY diagrams contributing to ${\cal O}$($\alpha_S$) to 
$h^o \to b \bar b$ decay}
\label{fig:fig1}
\end{center}
\end{figure} 

To one-loop and ${\cal O}$($\alpha_S$) there are two type of diagrams, shown 
in Fig.~\ref{fig:fig1}, that
contribute to $$\Delta_{SQCD}=\Delta_{SQCD}^{\rm loops}+
\Delta_{SQCD}^{\rm CT}\,.$$ 

The triangle diagram, with exchange of sbottoms and gluinos, 
contributes to $\Delta_{SQCD}^{\rm loops}$, whereas the bottom 
self-energy diagram contributes to the counter-terms part 
$\Delta_{SQCD}^{\rm CT}$. The exact results in the on-shell scheme are
summarized by,

\newpage
\begin{eqnarray}
          & {\Delta _{SQCD}^{\rm loops}}=  
          \frac{\alpha_s}{3 \pi} \Bigl\{ \Bigl[ \frac{2 M_Z^2}{m_b} 
        \frac{\cos\beta \sin(\alpha + \beta)}{\sin\alpha}
        (I_3^b \cos^2\theta_{\tilde b} - Q_b s^2_W \cos 2 \theta_{\tilde b})
        + 2 m_b +   {Y_b} \sin 2 \theta_{\tilde b} \Bigr]  \nonumber \\
        & 
        \times \Bigl[ m_b   {C_{11}} +
          {M_{\tilde g}} \sin 2 \theta_{\tilde b}   {C_0} \Bigr]
        (m_b^2,M_{h^o}^2,m_b^2;
         {M_{\tilde g}^2}, {M_{\tilde b_1}^2},
                {M_{\tilde b_1}^2})
                \nonumber \\
        & 
        + \Bigl[ \frac{2 M_Z^2}{m_b} 
        \frac{\cos\beta \sin(\alpha + \beta)}{\sin\alpha}
        (I_3^b \sin^2 \theta_{\tilde b} + 
        Q_b s^2_W \cos 2 \theta_{\tilde b})
        + 2 m_b -   {Y_b} \sin 2 \theta_{\tilde b} \Bigr] 
        \nonumber \\
        & 
        \times \Bigl[ m_b   {C_{11}} - 
         {M_{\tilde g}} \sin 2 \theta_{\tilde b}   {C_0} \Bigr]
        (m_b^2,M_{h^o}^2,m_b^2;
         {M_{\tilde g}^2}, {M_{\tilde b_2}^2},
         {M_{\tilde b_2}^2})
                \nonumber \\
        & 
        + \Bigl[ -\frac{M_Z^2}{m_b} 
        \frac{\cos\beta \sin(\alpha + \beta)}{\sin\alpha}
        (I_3^b - 2 Q_b s^2_W) \sin 2 \theta_{\tilde b}
        + Y_b \cos 2 \theta_{\tilde b} \Bigr]  \nonumber \\
        & 
        \times \Bigl[ 2 { M_{\tilde g}} \cos 2 
        \theta_{\tilde b}   {C_0} 
        (m_b^2,M_{h^o}^2,m_b^2; {M_{\tilde g}^2},
         {M_{\tilde b_1}^2}, {M_{\tilde b_2}^2})
        \Bigr] \Bigr\}\,, \nonumber
        \label{eqn:fulldgvertex}
\end{eqnarray}
\begin{eqnarray}
        & {\Delta_{SQCD}^{\rm CT}}
        = -  \frac{\alpha_s}{3\pi} \Bigl\{ \frac{ {M_{\tilde g}}}{m_b} 
        \sin 2 \theta_{\tilde b}
        \Bigl[  {B_0} (m_b^2; {M_{\tilde g}^2},
         {M_{\tilde b_1}^2})
        -   {B_0} (m_b^2; {M_{\tilde g}^2},
          {M_{\tilde b_2}^2}) \Bigr]  \nonumber \\
        & 
        - 2 m_b^2 \Bigl[  {B_1^{\prime}}(m_b^2;
         {M_{\tilde g}^2}, {M_{\tilde b_1}^2})
        +   {B_1^{\prime}}(m_b^2;
          {M_{\tilde g}^2}, {M_{\tilde b_2}^2}) \Bigr]
        \nonumber \\
        & 
        - 2 m_b  {M_{\tilde g}} \sin 2 \theta_{\tilde b}
        \Bigl[  {B_0^{\prime}}(m_b^2;
         {M_{\tilde g}^2}, {M_{\tilde b_1}^2})
        -  {B_0^{\prime}}(m_b^2; {M_{\tilde g}^2},
         {M_{\tilde b_2}^2}) \Bigr]
        \Bigr\} \,,\nonumber
        \label{eqn:fulldgwfct}
\end{eqnarray}
where we have used the standard notation for masses, couplings and mixing
angles, and  we have followed the definitions and conventions for the 
one-loop integrals $B_0$, $B_0'$, $B_1'$, $C_0$ and $C_{11}$ of 
ref.~\cite{Hollik}. Notice that 

\vspace{-0.4cm}

$$ Y_b \equiv  {A_b} + { \mu} \cot\alpha $$  
appears in
 $h^o \tilde b_R \tilde b_L$ coupling and, therefore, 
  will be responsible for sizeable contributions in the large $A_b$ and/or 
  $\mu$ limit. Our results agree with those of refs.~\cite{Dabelstein,cjs}.
   
 In order to compute  $\Delta_{SQCD}$ in the decoupling limit of very heavy 
 sbottoms and gluinos, we have considered the following simple assumption 
 for the MSSM parameters,  
  $$M_{SUSY} \sim M_{{\scriptscriptstyle {\tilde Q}}}
 \sim M_{{\scriptscriptstyle {\tilde D}}} \sim M_{\tilde g} 
 \sim \mu  \sim A_{b} \gg M_Z $$
 where the symbol '$\sim$' means 'of the order of' but not necessarily equal.
  We have performed a systematic expansion of the one-loop integrals and the
 mixing angle $\theta_{\tilde b}$ in inverse powers of the large SUSY mass
 parameters. The resulting formulas of these expansions can be found in 
 ref.~\cite{Haber:2001kq}. Thus, by defining 
 $$\tilde M_S^2 \equiv 
\frac{1}{2}(M_{\tilde b_1}^2 + M_{\tilde b_2}^2)\,\,,
   \,\,R \equiv \frac{M_{\tilde g}}{\tilde M_S}\,\,,\,\,
   X_b \equiv A_b-\mu \tan \beta  $$
   and including terms up to 
${\cal O}(M_{Z,h^o}^2/{\tilde M_S^2})$ in the expansion, 
we get the following result
for the maximal mixing case, $\theta_{\tilde b}\sim \pm 45 ^{\circ}$:
\begin{eqnarray}
        &  {\Delta_{SQCD}}
        = \frac{\alpha_s}{3\pi} 
        \left\{ \frac{-  {\mu M_{\tilde g}}}{  {\tilde M_S^2}}
        \left( \tan\beta + \cot\alpha \right)
        f_1(R) 
        - \frac{  {Y_b M_{\tilde g}} m_h^2}{12  {\tilde M_S^4}} f_4(R)
        \right.
        \nonumber \\
        &\qquad + \left. \frac{2}{3} \frac{M_Z^2}{ {\tilde M_S^2}}
        \frac{\cos\beta \sin(\alpha + \beta)}{\sin\alpha} 
        I_3^b
        \left( f_5(R) + \frac{ {M_{\tilde g} X_b}}{ {\tilde M_S^2}} 
        f_2(R) \right) + 
        {\cal{O}} \left( \frac{m_b^2}{ {\tilde M_S^2}} \right) \right\}
        \nonumber
        \label{eq:45degexpansion}
\end{eqnarray}
 where the functions $f_i(R)$ are defined in ref.~\cite{Haber:2001kq} and 
 have been normalized as $f_i(1)=1$.
 
 Notice that the first term is the dominant one in the
 limit of large $M_{SUSY}$ mass parameters and does not vanish 
 in the asymptotic limit of infinitely large 
 ${\tilde M_S}$, $M_{\tilde g}$ and $\mu$.  The second and third 
 terms are respectively  
 of ${\cal O}(M_{h^o}^2/{M^2_{SUSY}})$ and ${\cal O}(M_{Z}^2/{M^2_{SUSY}})$
  and vanish in the previous asymptotic limit. Therefore the first term gives
  a non-decoupling SUSY contribution to the $\Gamma(h^o \to b \bar b)$ partial
 width which can be of phenomenological interest. Moreover, since this term 
 is enhanced at large $\tan \beta$ it can provide important corrections to the
 branching ratio $BR(h^o \to b \bar b)$, even for a very heavy SUSY spectrum. The sign of these corrections 
 are fixed by the sign of $\mu M_{\tilde g}$. The previous result when expressed in
 terms of the $h^0$ effective coupling  to $b\bar b$ agrees with the
 result in ref.~\cite{cmwpheno} based on the zero external
 momentum approximation or, equivalently, the effective Lagrangian approach.
       
 From our previous result, we conclude that there is no decoupling of sbottoms
 and gluinos  in the limit of large SUSY mass parameters for fixed $M_A$. 
Notice that this result is at first sight surprising, since most 
numerical studies done so far on this subject indicate decoupling 
of heavy SUSY particles from SM physics\footnote{It should also be 
noticed that, strictly speaking, the {\it decoupling theorem}~\cite{ac}
is not applicable to the MSSM case, since it is a theory that incorporates 
the SM chiral fermions
and the SM electroweak spontaneous symmetry breaking. For a more detailed
 discussion on this, see ref.~\cite{dhp}}. 
  How do we then recover
 decoupling of the heavy MSSM spectra from the SM low energy physics? The
 answer to this question relies in the fact that in order to converge to SM
 predictions we need to consider not just a heavy SUSY spectra but also a heavy 
 Higgs sector. That is, besides large $M_{SUSY}$, the condition of large $M_A$ 
 is also needed. Thus, if $M_A \gg M_Z$ the light Higgs $h^0$ behaves as the SM
 Higgs boson, and the extra heavy Higgses $A$, $H^{\pm}$ and $H^0$ decouple. 
 The decoupling of SUSY particles and the extra Higgs bosons in 
 $\Delta_{SQCD}$ is seen explicitely 
 once the large $M_A$ limit of the mixing angle $\alpha$ is considered, 
  $${\cot\alpha} 
        =  {-\tan\beta} - 2 \frac{M_Z^2}{M_A^2} \tan\beta \cos 2\beta
        + {\mathcal{O}}\left(\frac{M_Z^4}{M_A^4}\right).$$
 By substituting this into our previous result we see that the non-decoupling
 terms cancel out and we get finally,
 \begin{eqnarray}
         {\Delta_{SQCD}} &
        = \frac{\alpha_s}{3\pi} 
        \left\{ \frac{2   {\mu M_{\tilde g}}} {  {\tilde M_S^2}}
        f_1(R)
            {\tan\beta} \cos 2\beta \frac{M_Z^2}{  {M_A^2}}
        - X_b\frac{  {M_{\tilde g}} m_{h^o}^2}{12 
         {\tilde M_S^4}} f_4(R)
        \right.
        \nonumber \\
        &\qquad + \left. \frac{2}{3} \frac{M_Z^2}{  {\tilde M_S^2}}
        \cos 2\beta  
        I_3^b
        \left( f_5(R) + \frac{  
        {M_{\tilde g}  {X_b}}}{  {\tilde M_S^2}} f_2(R) \right)
        + {\mathcal{O}} \left( \frac{m_b^2}{  {\tilde M_S^2}} \right) \right\}
        \nonumber
        \label{eq:45degexpansion}
\end{eqnarray}
 which clearly vanishes in the asymptotic limit of 
 $M_{\scriptscriptstyle {SUSY}}$ 
 and  $M_{\scriptscriptstyle {A}} \rightarrow \infty$.

 In conclusion, we get decoupling of the SQCD sector in 
 $h^0\rightarrow b\bar b$ decays, if and only if, both $M_{SUSY}$ and $M_A$ are 
 large.  In this limit, the dominant terms go as,
 $$\Delta_{SQCD}\sim C_1 \frac{M_Z^2}{M_A^2}+
 C_2\frac{M_{Z,h^0}^2}{M_{SUSY}^2},$$ 
 and, since both $C_1$ and $C_2$ are enhanced by $\tan\beta$, we expect
 this decoupling to be delayed for large $\tan\beta$ values. Last but not least,
 we see that the sign of $\Delta_{SQCD}$ is given by the sign of $\mu$ and 
 $M_{\tilde g}$. All these results are similar for the near zero mixing case,
 $\theta_{\tilde b}\sim 0^{\circ}$; for brevity we do not show these
 here (see ref.~\cite{Haber:2001kq}).
 
\begin{figure}[h]
\begin{center}
\epsfig{file=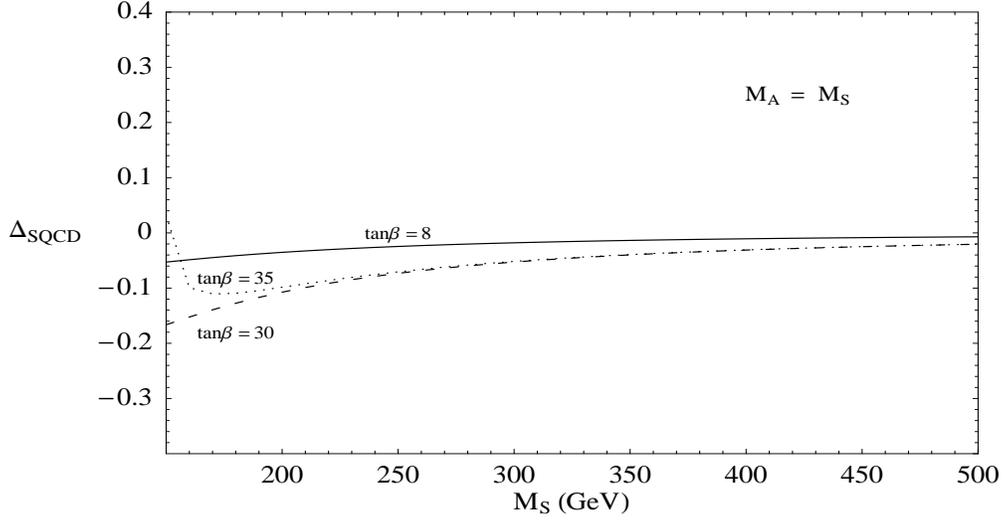,width=14cm,height=8.3cm}
\vspace{-1.2cm}
\caption[0]{Exact numerical results for $\Delta_{SQCD}$ in  
$h^0 \to b \bar b$ decay as a function of a
common MSSM scale $M_S$ and for several values of $\tan\beta$}
\label{fig:fig2}
\end{center}
\end{figure}
\vspace{-0.5cm}

Finally, in order to show this decoupling numerically, we have studied a simple 
example where there is just one relevant MSSM scale, $M_S$. More specifically, 
we have chosen,
$$ M_S =M_{\tilde Q}=M_{\tilde D}=\mu=A_b=M_{\tilde g}=M_{A},$$ 
which, in the limit $M_S \gg M_Z$, gives maximal mixing, 
$\theta_{\tilde b}\sim 45^{\circ}$. In Fig.~\ref{fig:fig2} we show the 
numerical results for the exact one-loop SQCD corrections, as a function of 
this common MSSM mass scale $M_S$, and for several values of $\tan\beta$. We
can see in this figure clearly the decoupling of $\Delta_{SQCD}$ with $M_S$.
This decoupling
goes as $1/M_S^2$, in agreement with our analytical result, and is delayed for
large $\tan\beta$ values. The typical size of this correction is 
$\Delta_{SQCD} \leq - 10 \%$ for $M_S\ge 250\,GeV$. Notice that the 
sign of $\Delta_{SQCD}$ here is
negative because of our choice of positive $\mu$ and $M_{\tilde g}$.     
 
%%%%%%%%%%%%%%%%%%%%%%%%%%%%%%%%%%%%%%%%%%%%
\section{Comparing the decoupling behaviour of the various MSSM 
sectors in $h^o \to b \bar b$ decay}
\begin{figure}[h]
\begin{center}
\epsfig{file=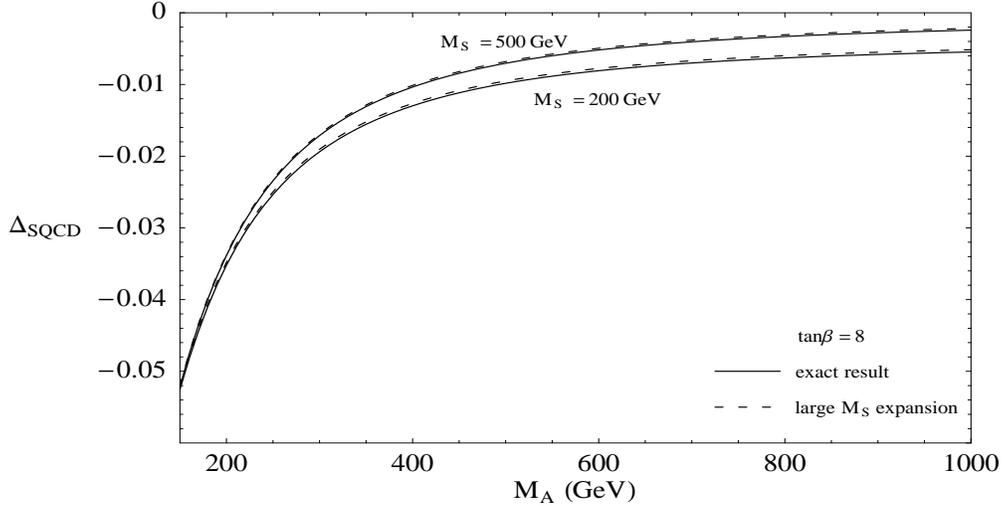,width=14cm,height=8.3cm} 
\vspace{-1.2cm}
\caption[0]{$\Delta_{SQCD}$ in $h^0 \to b \bar b$ decay as a function 
of $M_A$ for fixed $M_S$}
\label{fig:fig3}
\end{center}
\end{figure}

\begin{figure}[h]
\begin{center}
\epsfig{file=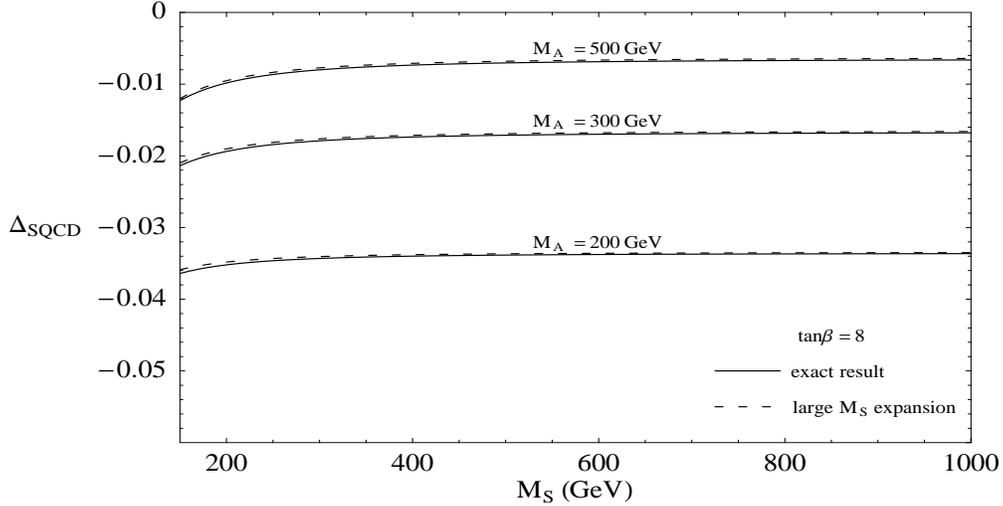,width=14cm,height=8.3cm}
\vspace{-1.2cm}
\caption[0]{$\Delta_{SQCD}$ in $h^0 \to b \bar b$ decay 
as a function of $M_S$ for fixed $M_A$}
\label{fig:fig4}
\end{center}
\end{figure}

In this section we study and compare the decoupling behaviour of 
the different MSSM 
sectors that are relevant in the one-loop SQCD corrections to 
$h^o \to b \bar b$ decay. As we have discussed in the previous section, these
are: the extra Higgs bosons, $H^0$, $A^0$, $H^{\pm}$, gluinos $\tilde g$ and
sbottoms $\tilde b_{1,2}$. As regard to the Higgs sector, we have seen
that there is no independent decoupling of these heavy 
$H^0$, $A^0$, $H^{\pm}$ Higgs bosons in $\Delta_{SQCD}$, unless the SQCD sector
is also considered heavy. To illustrate this, we have ploted in 
Fig.~\ref{fig:fig3} the numerical results of $\Delta_{SQCD}$ as a function of 
$M_A$ for several fixed values of the common SUSY scale  
$M_S =M_{\tilde Q}=M_{\tilde D}=\mu=A_b=M_{\tilde g}$ and for $\tan \beta=8$.
%\vspace{-0.5cm}
 The fact that  $\Delta_{SQCD}$ does not tend to zero for large $M_A$ but to a
 non-vanishing constant is a clear indication of a non-decoupling behaviour 
 with $M_A$ for fixed $M_S$. Similarly, we have shown that there is no independent 
 decoupling of the SQCD particles. That is, if we consider large values of the
 common SQCD scale $M_S$, while keeping $M_A$ fixed, $\Delta_{SQCD}$ approaches to a
 non-vanishing constant. This is illustrated clearly in Fig.~\ref{fig:fig4}.
%\vspace{-0.5cm}
 In these Figs.~\ref{fig:fig3} and~\ref{fig:fig4}
 we also show a comparison of the exact formula
 with our asymptotic expansion, valid for large $M_S$. We have seen that 
 the large 
 $M_S$ expansion provides a very good approximation to the exact result for
 most of the ($M_S,\tan \beta$) parameter space, except for sufficiently low $M_S$ 
 and large $\tan \beta$ values.
  
\begin{figure}[h]
\begin{center}
\epsfig{file=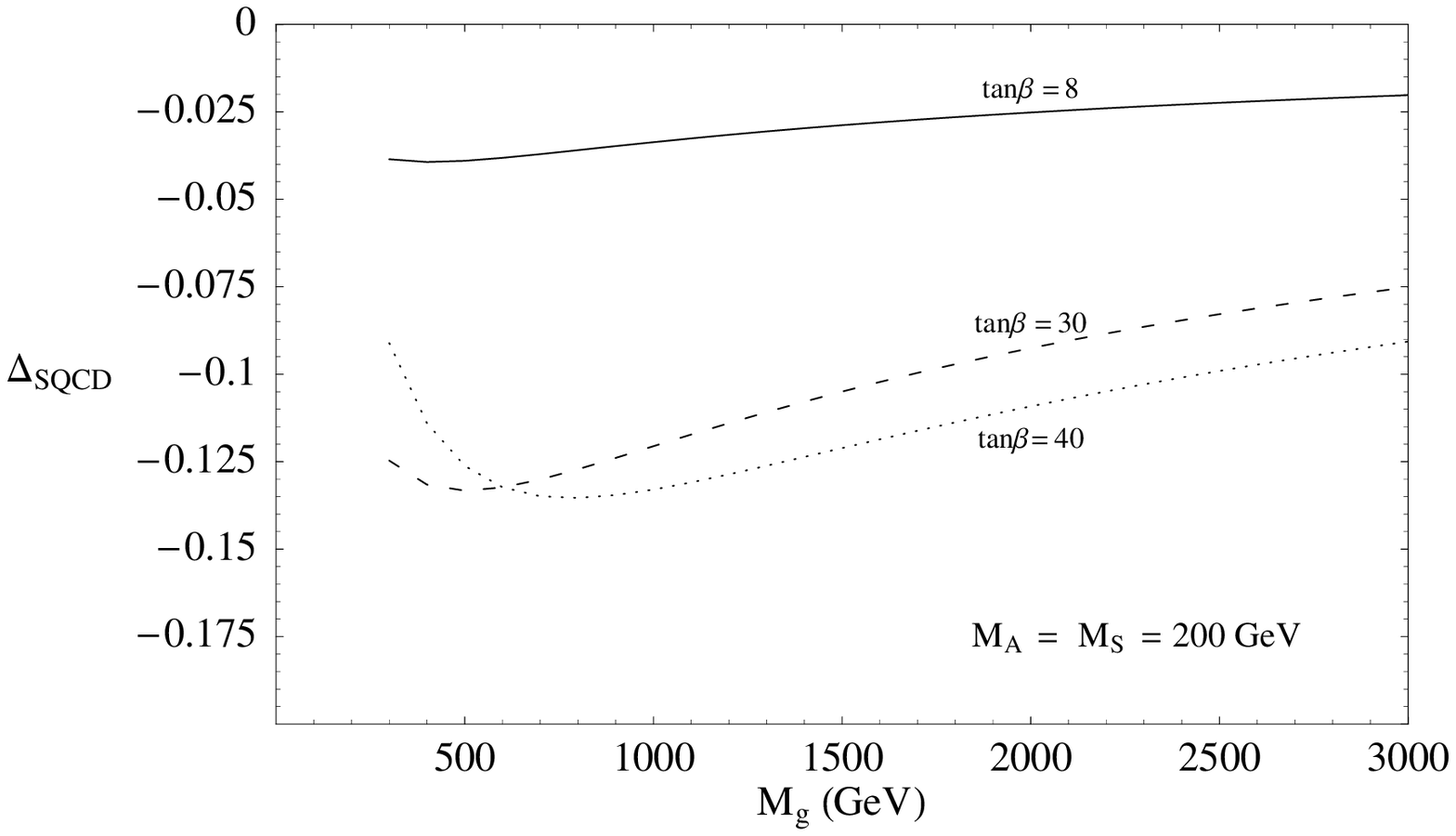,width=14cm,height=8.3cm} 
\vspace{-1.2cm}
\caption[0]{$\Delta_{SQCD}$ in $h^0 \to b \bar b$ decay 
as a function of $M_{\tilde g}$}
\label{fig:fig5}
\end{center}
\end{figure}

 Next, we consider the independent decoupling of gluinos. By performing 
 a second expansion in inverse powers of the gluino mass $M_{\tilde g}$,
 which is relevant in the heavy gluino limit 
 $M_{\tilde g} \gg \tilde M_S \sim \mu \sim A_b \gg M_Z$, we get:
 \begin{eqnarray}
          &{\Delta_{SQCD}}= 
        \frac{\alpha_s}{3\pi} 
        \Bigl\{ \frac{2 \mu}{  {M_{\tilde g}}}(  {\tan\beta} + \cot\alpha) 
        \left(1 -\log\left(\frac{M^2_{\tilde g}}{\tilde M_S^2} \right)\right)
        \nonumber \\
        &\qquad + \frac{2X_b}{  {M_{\tilde g}}} \frac{M_Z^2}{\tilde M_S^2}
        \frac{\cos\beta \sin(\alpha + \beta)}{\sin\alpha} I_3^b
        - \frac{Y_b}{3  {M_{\tilde g}}} \frac{m_h^2}{\tilde M_S^2}
        + {\mathcal{O}}\left(\frac{M^2}{  {M_{\tilde g}^2}}\right) \Bigr\}. 
        \nonumber
        \label{eq:heavygluinoexp}
\end{eqnarray}
 This yields a very slow decoupling with $M_{\tilde g}$, due to the
 logarithmic dependence, and agrees with the previous exact numerical
 results of ref.~\cite{cjs}. For illustration we show in Fig.~\ref{fig:fig5}
 our exact numerical results for $\Delta_{SQCD}$ as a function of 
 $M_{\tilde g}$, for fixed $M_A =M_S$ value and for several $\tan\beta$
 values.
%\vspace{-0.5cm}
This very slow decoupling with $M_{\tilde g}$ may have important
phenomenological consequences, in the large $\tan \beta$ regime,
because the SQCD correction can reach sizeable
values, even for large gluino masses. For instance, for $\tan\beta = 30$ and
$M_{\tilde g} = 1 TeV$  we get $\Delta_{SQCD} = -12 \%$, which is not a small
effect.

\begin{figure}[h]
\begin{center}
\resizebox{12cm}{!}{\rotatebox{270}{\includegraphics{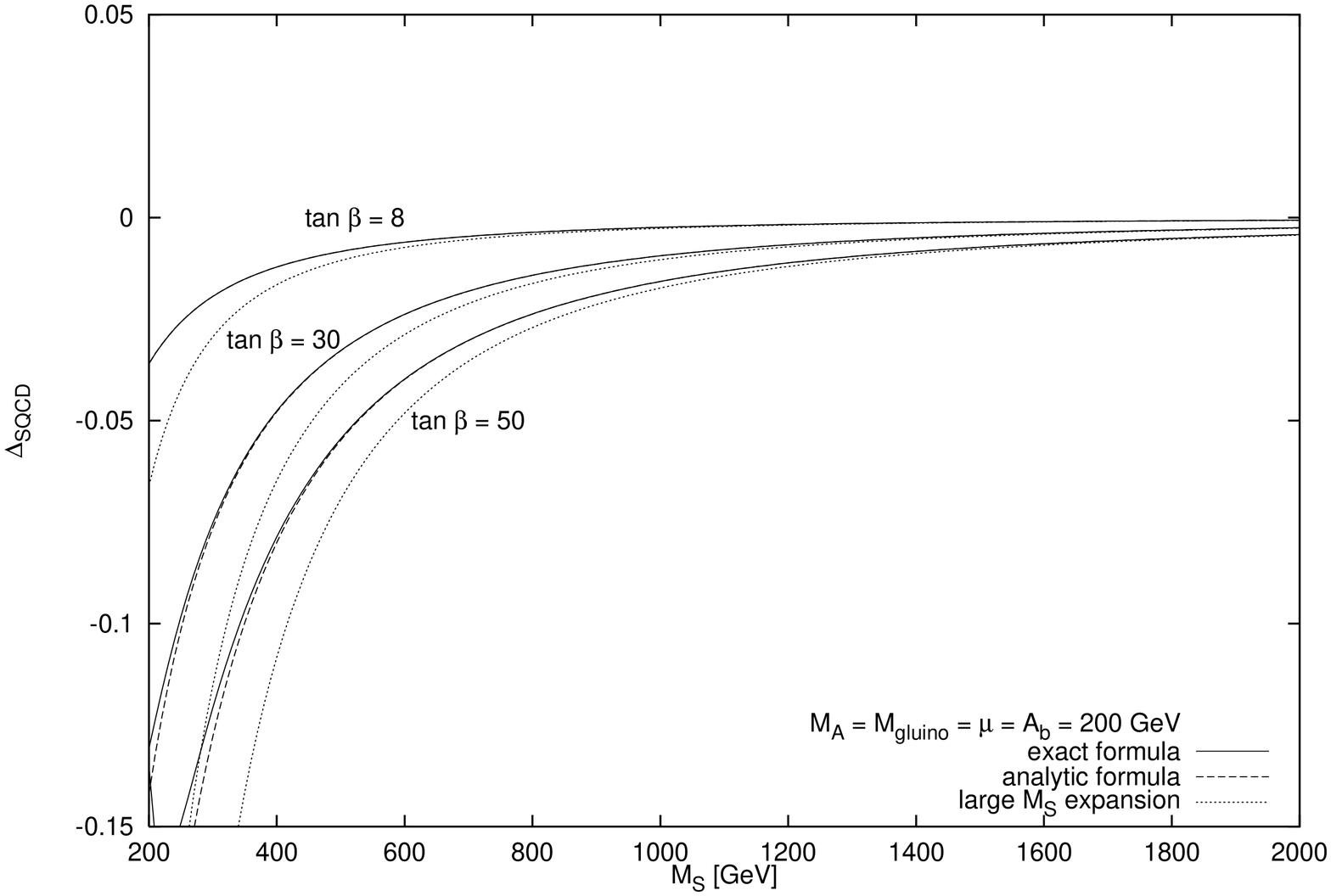}}}
\caption[0]{$\Delta_{SQCD}$ in $h^0 \to b \bar b$ decay 
as a function of ${\tilde M_S}$}
\label{fig:fig6}
\end{center}
\end{figure}

Finally, we have studied the independent decoupling of sbottoms. By performing
an expansion in inverse powers of the average sbottom mass ${\tilde M_S}$,
which is relevant in the heavy sbottoms limit,
$\tilde M_S \gg M_{\tilde g} \sim \mu \sim A_b \gg M_Z$ we find the following
result:  
\begin{eqnarray*}
        {\Delta_{SQCD}}
        = \frac{\alpha_s}{3\pi} 
        \left\{ \frac{-2 \mu M_{\tilde g}}{  {\tilde M_S^2}}
        (  {\tan\beta} + \cot\alpha)
        \right. 
% \nonumber \\
%         &\qquad \qquad \qquad 
\left.
        + \frac{M_Z^2}{  {\tilde M_S^2}}
        \frac{\cos\beta \sin(\alpha + \beta)}{\sin\alpha} I_3^b 
        + {\mathcal{O}}\left(\frac{m_b^2}{  {\tilde M_S^2}}
        \right) \right\}\nonumber
        \label{eq:largeSMexp}
\end{eqnarray*}
It shows a fast decoupling
behaviour as $\tilde M_S$ is taken large. This same behaviour is also manifest in our 
exact numerical results shown in Fig.~\ref{fig:fig6}   

%%%%%%%%%%%%%%%%%%%%%%%%%%%%%%%%%%%%%%%%%%%%%%%%%%%%%%%%%%%
\section{SUSY-QCD corrections to $H^+\rightarrow t\bar b$ in the decoupling 
limit}

In this section we study the SUSY-QCD corrections to the partial decay width 
$\Gamma(H^+\rightarrow t\bar b)$ at the one-loop level and to 
${\cal O}$$(\alpha_S)$. We will then analyze these corrections 
in the decoupling limit of large SUSY masses. We will present here just a short 
summary of the main numerical and analytical results, and refer the reader to 
ref.~\cite{hpt} for a more detailed study.

If all SUSY particles are heavy enough, $H^+$ decays dominantly into 
$t \bar b$ above the $t \bar b$ threshold. As in the case of 
$h^0\rightarrow b\bar b$, the dominant radiative corrections to 
$H^+\rightarrow t\bar b$ decay are the QCD corrections. At the one-loop 
level and to ${\cal O}$$(\alpha_S)$ the corresponding partial width can be 
written as,
$$\Gamma_1(H^+ \to t \bar b) \equiv \Gamma_0(H^+ \to t \bar b)
        (1 + 2   {\Delta_{QCD}} + 2   {\Delta_{SQCD}}),$$
where  $\Gamma_0(H^+ \to t \bar b)$ is the tree-level width, $\Delta_{QCD}$ 
is the correction from standard QCD, and $\Delta_{SQCD}$ is the correction from 
SUSY-QCD. The standard QCD corrections were computed in ref.~\cite{qcdhtb} and 
can be large ($+10\%$ to $-50\%$). The SUSY-QCD corrections were
        computed by using a diagrammatic approach in 
refs.~\cite{js,behkmy} and can be comparable or even larger than the 
standard QCD corrections in a large region of the SUSY parameter space. 

\begin{figure}[h]
\begin{center}
\epsfig{file=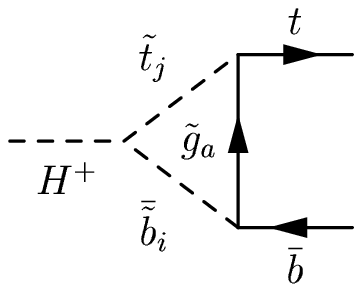}~~
\epsfig{file=AutoEn.ps}~~
\epsfig{file=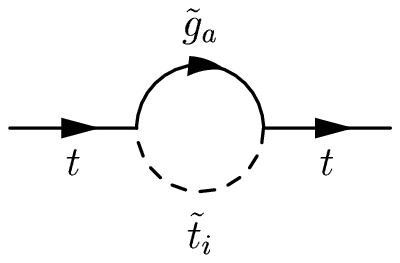}
\vspace{-1.2cm}
\caption[0]{One-loop SUSY diagrams contributing to ${\cal O}(\alpha_S)$
 to $H^+ \to t \bar b$ decay}
\end{center}
\label{fig:fig7}
\end{figure}

At the one-loop level and to ${\cal O}(\alpha_S) $ there are two type of diagrams that
contribute to $$\Delta_{SQCD}=\Delta_{SQCD}^{\rm loops}+
\Delta_{SQCD}^{\rm CT},$$ 
as shown in Fig.~\ref{fig:fig7}. 
%\vspace{-0.5cm}
The triangle diagram, with exchange of sbottoms, stops and gluinos, 
contributes to $\Delta_{SQCD}^{\rm loops}$, whereas the bottom and top
self-energy diagrams contribute to the counter-terms part 
$\Delta_{SQCD}^{\rm CT}$. The exact results in the on-shell scheme are
summarized by,

\begin{eqnarray}
&\Delta_{SQCD}^{loops} =  \frac{U_t}{D}\,H_t+
\frac{U_b}{D}\,H_b\,,\nonumber \\
&\Delta_{SQCD}^{CT} =  \frac{U_t}{D}\,\left(\frac{\delta m_t}{m_t}
        +\frac{1}{2}\,\delta Z_L^b+\frac{1}{2}\,\delta Z_R^t \,\right)+
 \frac{U_b}{D}\,\left( \frac{\delta m_b}{m_b}
        +\frac{1}{2}\,\delta Z_L^t+\frac{1}{2}\,\delta Z_R^b \, \right)\,,
\nonumber       
\label{eq:htbSQCD}
\end{eqnarray}
where,
\begin{eqnarray} 
D &=&
(M_{H^+}^2-m_t^2-m_b^2)\,(m_t^2\cot^2\beta+m_b^2\tan^2\beta)
-4m_t^2m_b^2\,,\nonumber\\ 
U_t &=&(M_{H^+}^2-m_t^2-m_b^2)\,m_t^2\cot^2\beta - 2m_t^2m_b^2\,,\nonumber\\ 
U_b &=&(M_{H^+}^2-m_t^2-m_b^2)\,m_b^2\tan^2\beta - 2m_t^2m_b^2\,,\nonumber\\
 H_t&=&- \frac{2\alpha _s}{3 \pi}\,\frac{G_{ab}^*}{m_t
\cot\beta} [m_t R^{(t)}_{1 b}R^{(b)*}_{1 a}(C_{11}-C_{12})+ 
m_b R^{(t)}_{2b}R^{(b)*}_{2 a}C_{12} \nonumber\\
& &+ M_{\tilde g} R^{(t)}_{2 b}R^{(b*)}_{1 a}C_0]
(m_t^2,m_{H^+}^2,M_{\tilde g}^2,M_{\tilde t_b}^2,M_{\tilde b_a}^2)\,,
\nonumber\\
H_b&=&- \frac{2\alpha _s}{3 \pi}\, \frac{G_{ab}^*}
{m_b \tan\beta} [m_t R^{(t)}_{2b}R^{(b)*}_{2 a}(C_{11}-C_{12})+ 
m_b R^{(t)}_{1 b}R^{(b)*}_{1 a}C_{12}
\nonumber\\
& &+ M_{\tilde g} R^{(t)}_{1 b}R^{(b)*}_{2 a}C_0] 
(m_t^2,m_{H^+}^2,M_{\tilde g}^2,M_{\tilde t_b}^2,M_{\tilde b_a}^2)\,,\nonumber
\label{eq:HLHR}
\end{eqnarray}
and the counter-terms are given in the on-shell scheme by,
\begin{eqnarray}
\frac{\delta m_{(t,b)}}{m_{(t,b)}}
        +\frac{1}{2}\,\delta Z_L^{(b,t)}+\frac{1}{2}\,\delta Z_R^{(t,b)} \,
        &=&
\Sigma^{(t,b)}_S(m_{(t,b)}^2)+ \frac{1}{2}\Sigma^{(t,b)}_L(m_{(t,b)}^2) -
\frac{1}{2}\Sigma^{(b,t)}_L(m_{(b,t)}^2)
\nonumber\\ 
- \frac{m_{t}^2}{2}
\left[\Sigma^{t'}_L(m_{t}^2) + \Sigma^{t'}_R(m_{t}^2)+ 2 \Sigma^{t'}_S(m_{t}^2)
\right]
&-& \frac{m_b^2}{2}
\left[\Sigma^{b'}_L(m_b^2) + \Sigma^{b'}_R(m_b^2)+ 2 \Sigma^{b'}_S(m_b^2)
\right]\nonumber 
\label{eq:CTselfE}
\end{eqnarray}
where,
\begin{eqnarray}
\Sigma^q_L(p^2)&=& - \frac{2\alpha_s }{3 \pi} 
\,|R^{(q)}_{1 a}|^2 B_1 (p^2,m^2_{\tilde{g}},m^2_{\tilde{q}_a})\,,\nonumber\\
\Sigma^q_R(p^2)&=& - \frac{2\alpha_s }{3 \pi}
\,|R^{(q)}_{2 a}|^2 B_1 (p^2,m^2_{\tilde{g}},m^2_{\tilde{q}_a})\,,\nonumber\\
\Sigma^q_S(p^2)&=& - \frac{2\alpha_s }{3 \pi}
\frac{m_{\tilde{g}}}{m_q}\, {\rm Re}
(\,R^{(q)}_{1 a}\,R^{(q)*}_{2 a}) B_0(p^2,m^2_{\tilde{g}},
m^2_{\tilde{q}_a})\,.\nonumber
\label{eq:selfs}
\end{eqnarray}
The $G_{ab}$ parametrize the $H^{+}\,\tilde b_a\,\tilde{t}_b$ couplings, and the
$R^{(q)}$ are the rotation matrices that relate the interaction-eigenstate
squarks to the mass-eigenstates. Their values in the MSSM can be found, for
instance, in ref.~\cite{hpt}. The above result agrees with the original
computation of refs.~\cite{js,behkmy}.

In order to compute $\Delta_{SQCD}$ in the decoupling limit of large SUSY 
masses, we have considered all
the soft-SUSY-breaking mass parameters and the $\mu$ parameter to be
of the same order (co\-llec\-tively denoted by $M_{SUSY}$) and much
heavier than the electroweak scale,  
$$ M_{SUSY} \sim M_{\tilde Q} \sim M_{\tilde U} \sim M_{\tilde D} 
\sim M_{\tilde g} \sim \mu \sim A_t \sim A_b \gg M_{EW},$$
and we have performed a systematic expansion in inverse powers of the large 
SUSY mass parameters. Notice that in this case it does not make sense to 
consider the alternative limit of large $M_A$, since this parameter provides 
the charged Higgs mass value and, therefore, it must be fixed. We have obtained
analytical expansions for  $\Delta_{SQCD}$ that include up to 
${\cal O}(M_{EW}^2/ {M^2_{SUSY}})$ corrections, for all the interesting 
limiting cases of maximal and minimal mixing, in both the stop and the sbottom
sectors. For brevity, we do not present here  the complete results, which 
can be
found in ref.~\cite{hpt}, and we just show the most relevant result, that is, 
the dominant term in this expansion for the particular choice 
of maximal mixing. 
Thus, for  
$  \theta_{\tilde b,\tilde t} \sim 45^o$  and 
$  \tilde M_S^2 \equiv 
\frac{1}{2}(M_{\tilde b_1}^2 + M_{\tilde b_2}^2)\equiv 
\frac{1}{2}(M_{\tilde t_1}^2 + M_{\tilde t_2}^2)$ we get:
\begin{eqnarray}
        & {\Delta_{SQCD}}
        = \frac{\alpha_s}{3\pi} 
        \left\{ \frac{-  {\mu M_{\tilde g}}}{  {\tilde M_S^2}}
        \left( \tan\beta + \cot\beta \right)
        f_1(R) + {\cal{O}} \left( \frac{M_{EW}^2}
        {  {\tilde M_S^2}} \right) \right\}
        \nonumber
        \label{eq:45degexpansion}
\end{eqnarray}
This leading term does not vanish in the heavy SUSY particle limit and,  
therefore, there is no decoupling of stops, sbottoms and gluinos in the
$\Gamma (H^+ \to t \bar b)$ decay width to one-loop level. This can be seen
clearly, for instance, for the simplest case of equal mass scales, 
$\mu=M_{\tilde g}=\tilde M_S$, where $f_1(R)=1$. This leading term, when
expressed in terms of an effective coupling of $H^+$ to $b\bar t$ is in
agreement with the previous results of refs.~\cite{cgnw,ehkmy} 
that were obtained in the zero external 
momentum 
approximation by using an effective Lagrangian approach. We see in this
result the enhancement of $\Delta_{SQCD}$ by $\tan\beta$, so that this
non-decoupling effect can be numerically important for large $\tan\beta$ 
values. As in the case of $h^0$, the sign of the SQCD correction is determined
by the sign of $M_{\tilde g}$ and $\mu$. We have obtained similar results for 
the case of minimal mixing, as can be seen in~\cite{hpt}.         

\begin{figure}[h]
\begin{center}
\epsfig{file=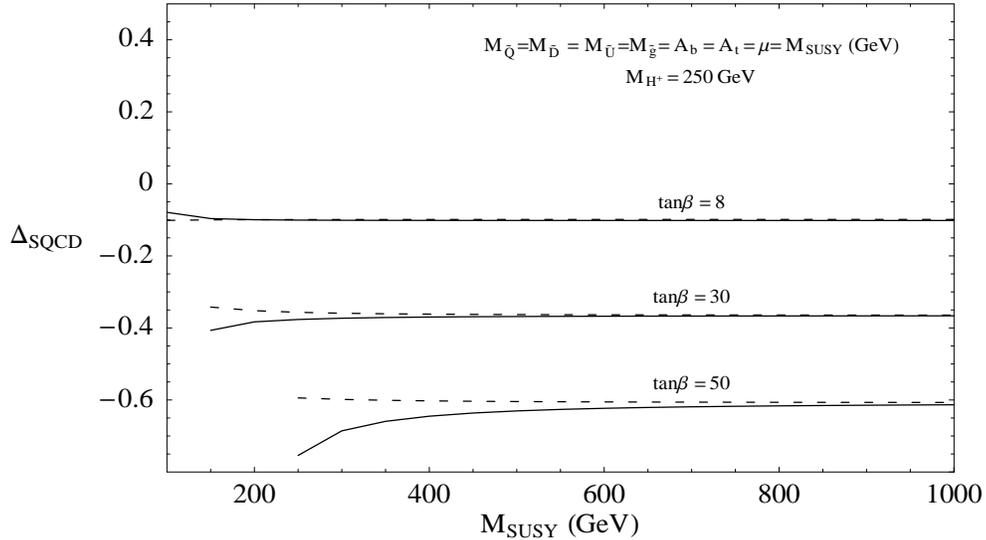,width=14cm,height=8.3cm} 
\end{center}
\vspace{-1.2cm}
\caption[0]{$\Delta_{SQCD}$ in $H^+ \rightarrow t\bar b$ decay as a function 
of the common SUSY scale $M_S$}
\label{fig:fig8}
\end{figure}

Finally, in order to illustrate this non-decoupling behaviour 
numerically, we present in Fig.~\ref{fig:fig8} the $\Delta_{SQCD}$ 
correction as a function of a common SUSY mass scale  
$M_S = M_{\tilde Q}= M_{\tilde U}=
 M_{\tilde D}= M_{\tilde g} = A_b = A_t = \mu$. The Higgs mass has been 
 fixed to $m_{H^+}=250\,GeV$, and several values of $\tan\beta$ have been
 considered. The fact that $\Delta_{SQCD}$ tends to a non-vanishing value
 for very large $M_S$ shows precisely this non-decoupling effect. 
 The correction is quite sizeable, even for a very 
 heavy SUSY spectrum. This is particularly noticeable for large $\tan\beta$.  
%\vspace{0.5cm}

\begin{figure}[h]
\begin{center}
\epsfig{file=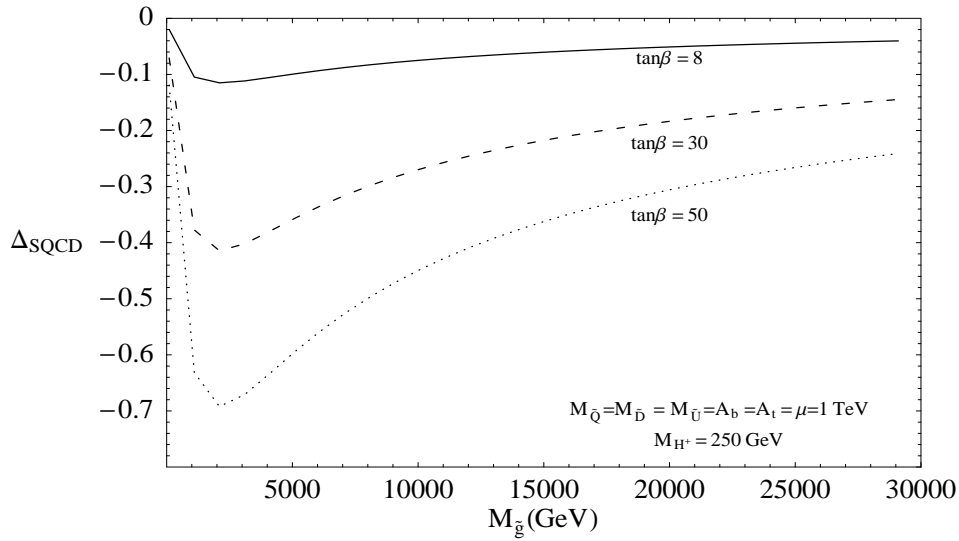,width=14cm,height=8.3cm} 
\end{center}
\vspace{-1.2cm}
\caption[0]{$\Delta_{SQCD}$ in $H^+ \rightarrow t\bar b$ decay as a function 
of $M_{\tilde g}$}
\label{fig:fig9}
\end{figure}

In addition, we have proved the independent decoupling of the gluinos and 
squarks whenever they are considered separately very heavy as
compared to the electroweak scale. Futhermore, the decoupling of gluinos 
is much slower than
the decoupling of squarks due again to the logarithmic dependence on the 
gluino mass. In Fig.~\ref{fig:fig9} we show the exact numerical results for 
$\Delta_{SQCD}$ as a function of the gluino mass and for 
$M_{\tilde Q}= M_{\tilde U}=
 M_{\tilde D}=  A_b = A_t = \mu= 1\,TeV$ and $m_{H^+}=250\,GeV$. We see clearly 
 the very slow decoupling of the correction with the gluino mass and notice
 the large size of $\Delta_{SQCD}$, specially for large $\tan\beta$.
 For instance, if  $\tan\beta = 30$ and 
 $M_{\tilde g} = 2$ TeV we get $\Delta_{SQCD} = -40 \%$. Notice that the size 
 can be so large that the validity of the perturbative expansion can be
 questionable. We refer the reader to refs.~\cite{cgnw,ehkmy} where this subject 
 is studied and
 some techniques of resummation for a better convergence of the series are
 proposed.
%%%%%%%%%%%%%%%%%%%%%%%%%%%%%%%%%%%%%%%%%%%%%%%%%%%%%%%%%%%%%%%%%%%%%%
\section{SUSY-QCD corrections to $ t \rightarrow W^+ b$  in the decoupling 
limit}
\begin{figure}
\begin{eqnarray*}
&&{\normalsize   \epsfig{file=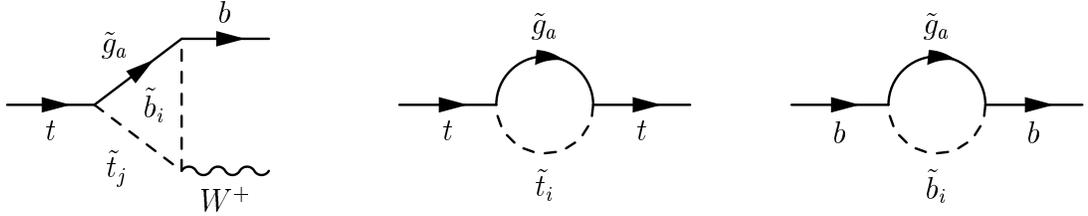}\ \ \epsfig{file=AutoEnt.ps}\ 
\ \epsfig{file=AutoEn.ps}}
\end{eqnarray*}
\vspace{-1.2cm}
\caption[0]{One-loop SUSY diagrams contributing to ${\cal O}(\alpha _S)$
in $ t \rightarrow W^+ b$ decay}
\label{fig:fig10}
\end{figure}

In this section we briefly comment on the SUSY-QCD corrections to $ t \rightarrow W^+ b$ 
at the one-loop level and to ${\cal O}(\alpha_S)$, and we study them in the
decoupling limit.  These radiative corrections 
were studied in the context of the MSSM in ref.~\cite{dhjjs}  
and are known to be important for some regions of the MSSM parameter space. 
 The standard QCD corrections are also known 
to be important and give a $\sim -10\%$ reduction in 
$\Gamma( t \rightarrow W^+ b)$~\cite{qcdtwb}. The Feynman diagrams
that contribute to the SQCD corrections are shown in 
Fig.~\ref{fig:fig10}.
%\vspace{-0.5cm}
The size of the SQCD corrections has been
estimated to range between $-5\%$ and $-10\%$ and are quite
insensitive to $\tan\beta$~\cite{dhjjs}. In contrast, the SUSY-Electroweak corrections that
range between $-1\%$ and $-10\%$ are known to grow with $\tan\beta$~\cite{ghjs}.
 
In order to analyze the decoupling limit in this observable we have chosen the 
simplest case with just one SUSY scale, $M_S$, which is considered very large 
as compared to the electroweak scale, $M_{EW}$, 
$$M_{\tilde Q}=M_{\tilde U}=
M_{\tilde D}=A_t=A_b=\mu=M_{\tilde g}=M_S \gg M_{EW}\,.$$

After performing an expansion of $\Delta_{SQCD}$ 
(we use here an analogous notation as in previous
sections) in inverse powers of $M_S$
we have obtained the following result for the dominant contribution,
\begin{eqnarray} 
\Delta_{SQCD} &= &-\frac{\alpha_s}{3\pi} \frac{m_t^2}{M_S^2}
\left(\frac{1}{6}+ 
\frac{1}{24}(1-\cot\beta)^2 +
\frac{1}{6} (1-\cot\beta)\right)+ {\cal{O}} \left( \frac{m_t m_W,m_W^2,
...}{M_S^2} \right)  \nonumber
\end{eqnarray} 
 From this result, we conclude that there is decoupling as $M_S$
 becomes large in the SQCD
 corrections to the 
 dominant top decay, $ t \rightarrow W^+ b$, and this decoupling which
 behaves as $(m_t^2/M_S^2)$ is not delayed. Indeed, we see in the previous equation that these corrections 
 are not enhanced by $\tan\beta$. Thus, we do not expect relevant 
 indirect signals from a heavy SUSY-QCD sector in this decay channel. 
%\newpage   
%%%%%%%%%%%%%%%%%%%%%%%%%%%%%%%%%%%%%%%%%%%%%%%%%%%%%%%%%%%%%%% 
\section{Conclusions}

In this work we have studied the one-loop SQCD corrections to the
partial widths of $h^0 \rightarrow b \bar b$, $H^+ \rightarrow t \bar b$
and $t \rightarrow W^+ b$ decays, in the limit of large SUSY masses.  In order
to understand analytically the behavior of the SQCD corrections in this
limit,
we have performed expansions of the one-loop partial widths that are valid for 
large values of the SUSY mass parameters compared
to the electroweak scale.
We have shown that for the SUSY mass parameters and $M_A$ large and all of the
same order, the SQCD corrections in $h^0 \rightarrow b \bar b$ decay 
decouple like the inverse
square of these mass parameters, and the one-loop partial width  
$\Gamma (h^0 \rightarrow b \bar b)$ tends to its SM value. In this case 
the effective low energy theory that one obtains after integrating out
all the heavy
non-standard modes of the MSSM is precisely the SM. However, if the mass 
parameters are not
all of the same size, then this behavior can be modified.  If $M_A$
is light, then the SQCD corrections to the 
$\Gamma (h^0 \rightarrow b \bar b)$ decay width
do not decouple in the limit of large SUSY mass parameters. We have also
presented and discussed here a similar non-decoupling SQCD correction to the  
$\Gamma (H^+ \rightarrow t \bar b)$ decay width. Given the closely related
structure of the various Higgs bosons couplings to the SM fermions,
one expects that similar SUSY non-decoupling effects will appear as well in 
other decay channels such as $H^0 \rightarrow b \bar b$, 
$A^0 \rightarrow b \bar b$ and $t \rightarrow H^+ b$. In the limit of
large SUSY mass parameters and light $M_A$ 
the effective low-energy theory, valid at the electroweak scale, should
contain
two full Higgs doublets with Higgs-fermion couplings of the general
type-III model~\cite{typeIII} that have no restrictions (other than those imposed by the SM
symmetries), since 
supersymmetry is not anymore a symmetry of this low-energy theory.
The particular values of the couplings in this low-energy effective Lagrangian 
are generated by integrating out all the heavy SUSY particles from the original 
MSSM Lagrangian, and they can be computed~\cite{dht}.  These non-decoupling SQCD 
corrections can be of phenomenological interest at present 
and future colliders. In particular they can provide some clues in the indirect 
search of a heavy SUSY spectrum at the LHC~\cite{chtt}.

We have also examined, in Higgs decays, some special cases in which there is a hierarchy among
the SUSY mass parameters.  In the case of maximal squark mixing with
$M_S$ large and the other SUSY mass parameters and $M_A$ of order a
common mass scale
$M$ (chosen such that $M_{EW}\ll M\ll M_S$), the SQCD
corrections decouple like $M^2/M_S^2$.
Second, we examined the case of a large gluino mass with the other SUSY
mass parameters of order a common mass scale $M$
(chosen such that $M_{EW}\ll M\ll M_{\tilde g}$).
In this case we found that the SQCD corrections
decouple more slowly, like
$(M/M_{\tilde g}) \log(M^2_{\tilde g}/M_S^2)$.

Finally we have studied the dominant decay of the top quark, in the decoupling
limit of large SUSY mass parameters, and we have found that the SQCD 
corrections decouple as ${\cal O}(\frac{m_t^2}{M_S^2})$. It will be,
therefore, very difficult to look for indirect heavy SUSY signals in this
channel.

%%%%%%%%%%%%%%%%%%%%%%%%%%%%%%%%%%%%%%%%%%%%%%%%%%%%%%
\Acknowledgments
M.J.H. wish to thank Howard E. Haber and the Organizing Committee for
the kind invitation to give a talk at this Symposium and for the enjoyable atmosphere at
this interesting conference. S.P. acknowledges H.E. Haber for the invitation 
to attend this Symposium. 
This work has been supported in part by
the Spanish Ministerio de Educacion y Cultura under project CICYT
AEN97-1678.  S.P. has been partially supported by the
 U.S. Department of Energy under contract DE-FG03-92ER40689 and by
Ramon Areces Foundation.
S.R. has been partially supported by the European
Union through contract ERBFMBICT972474. H.E.H. is supported in part
by the U.S. Department of Energy under contract DE-FG03-92ER40689.
Fermilab is operated by Universities Research Association Inc.\
under contract no.~DE-AC02-76CH03000 with the U.S. Department of
Energy.

%%%%%%%%%%%%%%%%%%%%%%%%%%%%%%%%%%%%%%%%%%%%%%%%%%%%%

\end{document}

%% file: econf2.tex
\def\beq{\begin{equation}}
\def\eeq#1{\label{#1}\end{equation}}
\def\eeqn{\end{equation}}

\newenvironment{Eqnarray}%
   {\arraycolsep 0.14em\begin{eqnarray}}{\end{eqnarray}}
\def\beqa{\begin{Eqnarray}}
\def\eeqa#1{\label{#1}\end{Eqnarray}}
\def\eeqan{\end{Eqnarray}}

\let\bar=\overbar

\def\Dslash{\not{\hbox{\kern-4pt $D$}}}
\def\dslash{\not{\hbox{\kern-2pt $\del$}}}

\def\msb{{\bar{\ssstyle M \kern -1pt S}}}

\def\lsim{\mathrel{\raise.3ex\hbox{$<$\kern-.75em\lower1ex\hbox{$\sim$}}}}
\def\gsim{\mathrel{\raise.3ex\hbox{$>$\kern-.75em\lower1ex\hbox{$\sim$}}}}